\documentstyle[aps%
, preprint
, floats
, epsf]{revtex}

\begin{document}
\newcommand{\myfoot}[2]{\footnote{#2}}
\newcommand{\bibfoot}[2]{}
\draft
\preprint{\hfill\parbox{3.5cm}{KUL-TF-94/21 \\ hep-th/9602066}}
\tightenlines
\title{
Classical Tunneling  from the Lorentz-Dirac Equation
}
\author{Frederik Denef, Joris Raeymaekers, Urban M. Studer and
Walter Troost}
\address{
Instituut voor Theoretische Fysica,
Katholieke Universiteit Leuven,
Celestijnenlaan 200D, B-3001 Leuven, Belgium
}
\date{\today}
\maketitle
\begin{abstract}
The classical equation of motion of a charged point particle,
including its radiation reaction, implies tunneling.
For nonrelativistic electrons and a square barrier, the solution is 
elementary and explicit. We show the persistance of the solution
for smoother potentials. 
For a large range of initial velocities, initial conditions
may leave a (discrete) ambiguity on the resulting motion.
\end{abstract}

\narrowtext
In classical relativistic electrodynamics the motion of an electrically
charged point particle (without further structure; we will call
it an 'electron' for brevity.) is governed
by the Lorentz-Dirac (LD)
equation \cite{Dirac}.
For general reference, see \cite{general},
and  especially \cite{Rohrlich}.
It includes the effect of the back-reaction
of the (retarded) field generated by the electron during
its past on
its own motion.  The radiation reaction is taken into account through a
renormalisation procedure in which the bare mass of the electron and the
electromagnetic self-energy combine to the physical inertial mass. 
We will show that this classical equation exhibits tunneling.

There
exists a variety of methods
\cite{Dirac,Rohrlich,FeynWheel,Coleman,Barut,TiraBra,Teitelboim}
to derive this equation, the most comprehensive ones being
based on energy and momentum conservation. The LD equation emerges as the
unique result, if one assumes that, except its charge,
the electron possesses no other
attributes like dipole or higher multipole moments.
It reads
\begin{equation}
\stackrel{..}{z}^\mu=\frac{e}{m}F^{\mu\nu}\stackrel{.}{z}_\nu
  +\tau (\frac{\stackrel{..}{z}^2\stackrel{.}{z}^\mu}{c^2}
         +\stackrel{...}{z}^\mu)
\label{LDE}
\end{equation}
where $z^\mu$ is the position of the electron (and the singularity of its
electromagnetic field), dots denote derivatives with respect to proper
time, and
$\tau=\frac{2}{3}\frac{e^2}{4\pi \epsilon_0 m c^3}\simeq 0.62\ 10^{-23}\ \rm s$
is the 'pre-acceleration time'.
The Lorentz force term $F^{\mu\nu}$ only includes
the electromagnetic field generated by external sources. 
The last two terms in eq.(\ref{LDE}) are due to the radiation reaction.
The first is (minus) the derivative of the radiated four momentum, and the 
second, the Schott term,  
can be combined with the left hand side into the time rate of change of
$p_\mu=m (\stackrel{.}{z}_\mu -\tau \stackrel{..}{z}_\mu )$, the "bound
momentum". That momentum can consistently be interpreted as  the total
momentum of the electron together with its bound field\cite{Teitelboim}.
We will simply call it the momentum in the sequel. Asymptotically, when
the acceleration ceases, it is equal to the usual momentum.

To set in perspective the remarkable fact 
that this {\em classical} equation allows
the electron to tunnel through potential barriers
that are narrow enough to be crossed in a proper time $\tau$,
we first discuss some general features of eq.(\ref{LDE}).
Although the Lorentz-Dirac equation describes the radiation reaction in a
satisfactory way, and has been used in a variety of circumstances -- the
most extreme ones being the astrophysical applications \cite{AF} -- it
also exhibits some features that have raised eyebrows. For this third
order equation, the initial value problem demands a specification of
initial position, velocity {\em and acceleration}. The solution is then
unique, but in general unphysical, describing a 'runaway', i.e. a motion
with an exponentially increasing velocity, even in a force-free spatial
region. The specification of initial acceleration has therefore
traditionally been replaced by an {\em asymptotic} condition\myfoot{f3}{or,
perhaps more physically, by formulating the LD equation as an
integro-differential equation.} stating that the (unobserved) runaways
are rejected. From the mathematical point of view, the first question to
be answered is then, whether a satisfactory solution still exists for
reasonable initial data, and whether it is unique. This last point is
"one of the most important unsolved problems of the theory"
\cite{Rohrlichcit}, and remained unresolved, even though some isolated
cases have been found \cite{Huschilt} of initial data (for position and
velocity) that allow more than one solution. 
The LD-equation also gives rise to a new physical phenomenon:
pre-acceleration.
Indeed, the effect of the asymptotic boundary condition in the future
makes itself felt at early times, implying that an electron approaching a
sharply delineated region in space where a force field is present,
actually starts accelerating {\em before} it reaches the force field.
This is sometimes viewed upon as an undesirable feature.
A number of
alternatives to the LD equation have been (re-)proposed over the years
\cite{FOC}, but either run into difficulties
(mainly with energy conservation), or necessitate additional structure.
Although
some of them may be valid as models for an electron that has an extended
structure, we will not consider these alternatives further. 

In view of the considerable attention that the LD equation, and its
solutions, have generated, it is remarkable that the class of
solutions that we are about to describe,
with  surprising physical implications, has gone unnoticed so far.
They describe  {\em tunneling}.
Explicit examples can be found  by 
considering the familiar  setting from quantum mechanics:
a one dimensional problem, where an electron impinges on a region with a
rectangular potential energy barrier.
For our purposes the nonrelativistic approximation (NRA) will be sufficient.
Let us denote the
electrostatic field as $F=-dV/dx$. Choosing
units such that the electron mass, the velocity of light, and the
characteristic time $\tau$ are equal to unity,
the  equation becomes, with $v=\stackrel{.}{x}$,
\begin{equation}
\stackrel{.}{p}=\stackrel{.}{v}-\stackrel{..}{v}=F.\label{NRLDE}
\end{equation}
The electron  only experiences a force when
crossing the boundaries of the regions of constant potential.
The solutions in the separate force-free regions, which are easy to write
down explicitly, are connected using the following matching condition%
\myfoot{4}{This can be checked using the formal equation
$\stackrel{.}{p}=\Delta V \cdot \delta(x)$. The validity of the rectangular
barrier idealization will be discussed shortly.}
on the momentum, or the acceleration, the position and  the velocity being
continuous:
\begin{equation}
\Delta p = -\Delta \dot{v} =- \Delta V /v \ .\label{match acc}
\end{equation}
The asymptotic condition is most easily implemented by solving the
matching conditions backward in time, putting $\stackrel{.}{v}_f=0$
(a procedure that is also expedient when
numerically integrating the equation). In the non-relativistic
approximation eq.(\ref{NRLDE}), this results in the following set
of equations relating the initial and final velocities to the
time $T$  spent in the barrier region of  width $w$ and height $V$:
\begin{eqnarray}
w&=& v_f T- \frac{V}{v_f}(e^{-T}-1+T) \ ,\nonumber\\
v_i&=& v_f -\frac{V}{v_f}+\frac{V}{v_f-\frac{V}{v_f}(1-e^{-T})}\ .
\end{eqnarray}
Although the analysis of these equations 
in general is not very difficult, it becomes
particularly simple when the final electron energy is equal to half the
barrier height, $V=v_f^2$. An explicit example is, for $V=144,w=3$:
\begin{equation}
\begin{array}{rcrcl}
x=&-7 (e^t-1)+16 t\mbox{    }    &   &t&<0     \ ,  \\
  & 9 (e^t -1)                       & 0<&t&<T \ ,  \\
  & 3+12 (t-T)                   & T<&t&       \ ,
\end{array}        \label{spesol}
\end{equation}
with $T=\log 4/3$. The matching condition eq.(\ref{match acc}) implies jumps
of $16$ and $-12$ units in the acceleration at $t=0$ and $t=T$.
Tunneling occurs, the initial energy is equal to 128, 
a fraction 1/9 below the barrier height.

Whereas the details of the example above are of course special,
the tunneling phenomenon is actually quite generic. A decisive parameter
is the width of the potential. If the electron can cross the barrier
within a time of order 1, i.e. the pre-acceleration time $\tau$, tunneling
occurs. 
To understand this it suffices to follow the bound
momentum while the electron crosses the barrier. Its value is piecewise
constant, and  in the regions outside the
barrier equal to its asymptotic value.  Under  the  barrier  itself
the value is $v_f-V/v_f$, which may be in the same direction
{\em or opposite} to the
velocity. The  example  above  is  special,  in that the intermediate
momentum  vanishes.
Without external force, if the velocity is directed opppositely, 
it will quickly turn in a time of order 
$\tau$ in the same direction as the (conserved) momentum.
If  the width is small enough, the electron
has in the meantime reached  the opposite side of the barrier.
The  smaller  the  width,  the  wider  the  range  of  initial
velocities  for  which  the  electron  will tunnel through. 

The explicit solution given above can be used to illustrate
another remarkable property
of the LD equation (together with the asymptotic condition),
viz. the failure of initial data to determine the
solution uniquely. If we take a barrier 
extending from $x=3$ to some value $x<-9$,
and specify $x=-9$ and zero velocity in the infinite past, the
second line of eq.(\ref{spesol}), extended to  negative times,
together with the last line, constitute a solution that is an alternative
to the trivial one with constant $x$.
This  nonuniqueness  is  also  mentioned  in \cite{Huschilt}.
We do not regard this as a sufficient answer to the initial condition 
question of Rohrlich \cite{Rohrlichcit} cited above.
It is analogous to the ambiguity present already in
Newtonian mechanics, when specifying, in the infinite past, a zero velocity
at the top of a mountain: it
is an  isolated special  case, and an  infinitesimally  small
initial velocity eliminates the  stationary   solution.
Of considerably more interest is the fact that
rectangular  barrier  crossing,  whether by tunneling through or by passing
over  it,  very  often  exhibits a much more generic non-uniqueness%
\myfoot{f5}{A similar phenomenon has been noticed also in \cite{Gull}.
However, that study has some problems, see below.}.
For values of the initial velocity sufficiently large to cross the barrier,
there is in general more than one distinct solution,
typically one where almost all energy
is radiated away and up to five (see later)
where a smaller radiation loss occurs.
This shows that the non-uniqueness is a common feature of the LD equation:
given the no-runaway condition, it is generically still
insufficient to specify position and momentum at some initial time. This
is true not only for asymptotic initial values, but also for initial
values at finite times.

It  is  important  to  realize  that  the  features  discussed above do
{\em not} depend  on  the non-relativistic approximation involved.
There is a simple scaling property
of  the  NRA  that  leaves  this  equation  invariant,  viz.  rescaling all
lengths,  velocities  and  accelerations  with  a  common  factor,  and the
barrier height  with  its  square. This implies that we can always rescale
such  that  only  small  velocities are involved, and our discussion applies.
The  existence  of  the  tunneling  solutions is therefore beyond doubt.

What  is  less  clear,  is the possible role played by the sharpness of the
potential  step.  Indeed,  for  the discontinuous potential step, it turns
out that the  problem  as  formulated above has {\em no} solution for small
incident velocities.
This was noticed, for a single potential step, in \cite{Gull}.
It was stated there that no mathematical difficulties are associated
with the idealization of a sharp step. Physically, however, this absence of
solutions for a given initial velocity range is
probably unacceptable. To  investigate this point, and to
ascertain  at  the  same  time that the tunneling feature is not correlated
to  this  unphysical behavior, we have repeated the analysis when the
electron climbs a ramp with a finite slope.
We take the force $F$ to be a constant over a
region  of  width $\epsilon$, with%
\myfoot{f6}{We will always
consider  a  fixed  height  $V$  when  we  consider  the  small  $\epsilon$
limit.} $V=\epsilon F$.
At  the  moment we consider only a single ramp, we will come back
to  the  tunneling situation  later. Again, in each of the regions of
constant  force,  the  LD  equation can be solved exactly (NRA) in terms of
elementary  functions, for instance in the sloping region 
$ x= -F (e^t-1-t-t^2/2) +v_f t$. Then we again investigate
and  the  matching  conditions. We use the
same  backward-in-time method as before.

\begin{figure}
\epsffile[24 349 318 790]{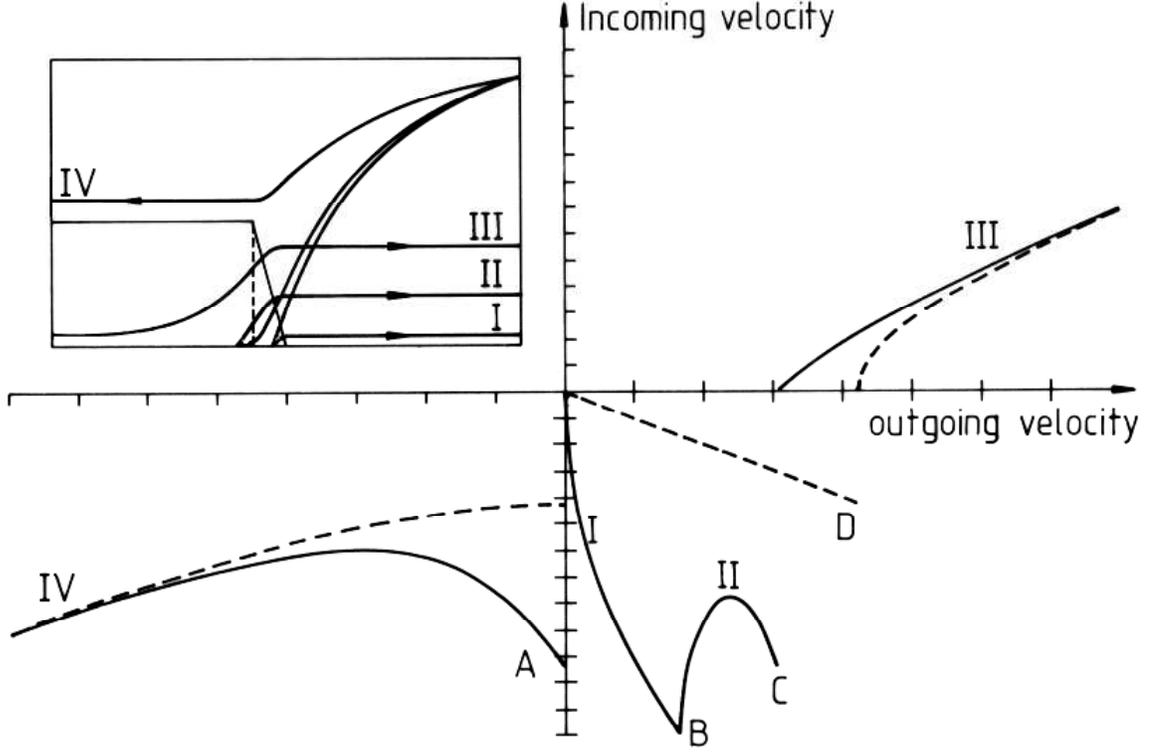}
\vspace{-5cm}
\caption{Plot of the initial velocity vs. the final velocity
for the solution of the Lorentz-Dirac equation in a linearly rising 
step potential (note the difference in scale). 
The dotted lines leave out the radiation reaction.
The inset shows the potential and the kinetic energy.
The four types of motion are discussed in the text. For the plots,
a step height $V=9$ was used, and a slope width $\epsilon=0.5$.
}
\label{rampfiguur}
\end{figure}

In figure \ref{rampfiguur} (see the inset),
we have plotted the kinetic energy
as a function of the position for typical cases. The four different 
possibilities that arise are listed in table \ref{tabel}.

\begin{table}                                                         
\caption{                                                             
Overview of the four different types of motion, described in the text.
\label{tabel}}                                                        
\begin{tabular}{lclc}                                                 
case & $v_i$ &Turning point         & $v_f$\\                         
\hline                                                                
I    & $-$     &inside sloping region & $+$ \\                        
II   & $-$     &under plateau         & $+$ \\                        
III  & $+$     &                      & $+$ \\                        
IV   & $-$     &                      & $-$ \\                        
\end{tabular}                                                         
\end{table}

Figure \ref{rampfiguur} also shows a plot of the initial versus the 
final velocity. The following remarks can be made.

Case~III: This motion agrees
fully with intuition. For a gentle
slope   ($\epsilon  \rightarrow  \infty$)  the  radiation  energy  loss  is
negligible, so that the final velocity is always larger than $\sqrt{2 V}$;
for  a  steep slope  ($\epsilon \rightarrow 0$) with the same height,
and small initial velocity, half the energy is radiated away due to the
larger acceleration, and the final velocity approaches $\sqrt{V}$. 

Case~IV: Here it is of
course necessary that the electron has enough energy ($>V$) to overcome
the barrier, and if the slope is gentle this is sufficient. For a steep
slope the minimum initial energy is $2V$. Note however the  surprising
feature  that the solution is not fixed by the initial
velocity alone: for a range of initial velocities larger than the minimum
required to overcome the barrier, there are actually two different
solutions. This range becomes larger as the barrier gets steeper, as
$6^{1/3} V^{2/3} \epsilon ^{-1/3}$ (point $A$ in the figure).
So we see now that nonuniqueness of solutions indeed persists when the
potential step is replaced with a slope, and in fact we have 
also checked it numerically for a completely smooth
ramp (a hyperbolic tangent). 

Cases~I and~II: Both
represent  reflections. The branch
starting at the origin corresponds to case~I, with $\epsilon \rightarrow 0$ 
limiting behavior $v_f
\sim v_{i} ^2 \epsilon /6V$. The point $B$ where case~II takes over is
located at $v_i \sim 3^{5/6} V^{2/3} \epsilon ^{-1/3} $. This shows that
the behavior for a steep ramp is quite subtle.
If one investigates only the (formal) limiting
equation without taking this into account, one is likely to miss branch I %
\myfoot{f8}{The numerical study in \cite{Gull} exhibits this
problem. We therefore reject the conclusion reached there, that for small
initial velocities no solutions would exist at all.}
although it is
clearly a physically correct possibility, and in fact, for small
velocities, this solution is unique. In the small $\epsilon$ limit, this
branch tends to the vertical axis. The type~II branch on the other hand
has a smoother limit, and  reduces to the straightforward
solution for infinite slope, as obtained from the matching condition
eq.(\ref{match acc}). There is no type~II solution beyond point $C$ in the
figure: electrons with a larger final velocity necessarily originate from
the  plateau, and are shown on branch~III.
For a very gentle slope, the whole compound curve~I~and~II,
will approach the line $v_{i}=-v_f$ representing no radiation loss,
while both points B and C converge to D ($v_f=\sqrt{2V}$).
For a very steep slope, both
points~$B$~and~$C$ move towards infinite initial velocities as $\epsilon
^{-1/3}$ (with a fixed ratio $\sqrt{3}/\sqrt[3]{2}$), and limiting final
velocities equal to $0$ and $\sqrt{V}$ respectively. 

Thus, when a high
velocity electron meets a very steep well, there is an amazing
variety of different possible outcomes. It may lose some energy and
travel on (curve~IV, left branch), just barely make it up the hill
(curve~IV, right branch) having lost most of its energy in radiation, or
be reflected with a choice of {\em three} different velocities!

It is clear that analogous results hold for a barrier with finite width, 
and that tunneling solutions will persist for $\epsilon \not= 0$. 
Furthermore,
we have checked numerically that the qualitative behavior discussed
above is unaltered when using relativistic kinematics
(with $\gamma v$ instead of the velocity as a parameter),
and also in the case of
an analytic (but rapidly varying) potential. 
The tunneling solutions obtained for the rectangular barrier are limiting
solutions of those for a smoothened barrier (e.g. the potential of
closely packed point charges). Thus we conclude that both the tunneling
phenomenon and the nonuniqueness of physical solutions are general
properties of the LD equation, and not artefacts due to unphysical
properties of the potential or the non-relativistic approximation.

The key to the physical  understanding 
of these phenomena is the use of the {\em bound} momentum
$p_\mu$ introduced in \cite{Teitelboim}.
Apart from the radiation loss
(the second term on the r.h.s. of eq.(\ref{LDE})), 
its rate of change is given by 
the external force exerted on the electron.
If the acceleration is not too large the difference with the "bare" 
momentum is just a mass renormalization, but when the electron velocity 
changes rapidly the accompanying self-field needs some time to adjust
to the new velocity (the updating is limited by the finite light speed), 
and $p_\mu$ is no longer simply proportional to $\stackrel{.}{z}_\mu$.

When the electron attacks a steep  slope, the bound momentum has to 
decrease very rapidly. The electron can simply decelerate and bounce back, 
but if the potential is narrow enough there is a second possibility: the 
electron can make a ``jump'', i.e. a short acceleration, during which the 
bound momentum {\em decreases}. Because of its {\em negative} bare mass, 
the bare electron gives a negative contribution to the bound momentum, 
which cannot  immediately be compensated entirely by the accompanying 
Coulomb field. When the acceleration ceases, the Coulomb field catches up 
and the bound momentum increases again, as it should once it reaches the 
downward slope at the other side of the barrier. In this way tunneling 
can take place. Note that the kinetic energy $p_0-m$ becomes negative in 
the classically forbidden region.

The essential feature of tunneling is that the 
crossing has to take place in proper times of the order of the
pre-acceleration time. For larger widths this could be obtained by 
considering very high speed electrons, which would effectively see a 
Lorentz-contracted barrier. It is theoretically not  difficult to construct 
arrangements of individual charges that might show the tunneling 
phenomenon for  very fast electrons%
\myfoot{f10}{In such cases one should also expect to have to
take into account quantum effects.}.
Whereas in some astrophysical applications (for example the
motion of charged particles in fields produced by pulsars\cite{AF})
there is a combination of fast electron motion with strong fields that 
necessitates the use of the Lorentz-Dirac equation, it is not clear 
whether they would provide a testing ground for the tunneling phenomenon 
described in this paper. 

A rough estimate indicates that for phenomena taking place in
times of order $\tau$ quantum considerations should enter.
Since quantum electrodynamics is arguably the most successful physical 
theory known, it would be interesting to investigate its relation to the 
tunneling phenomenon discussed in the present paper, and more generally to 
the Lorentz-Dirac equation. This is outside the scope of the present 
letter.

\acknowledgments

F.D. and W.T acknowledge the financial support of 
the N.F.W.O. (Belgium), and  U.S. the support of the Research Council of the 
K.U.Leuven.



\end{document}